\documentclass[conference,10pt]{IEEEtran}

\usepackage{fancyhdr}

\usepackage{caption}
\usepackage{hhline}
\usepackage{enumitem}
\usepackage[hidelinks]{hyperref}
\usepackage{cite}
\usepackage{amsmath,amssymb,amsfonts}
\usepackage[linesnumbered,ruled,vlined]{algorithm2e}
\usepackage{algpseudocode}
\usepackage{graphicx}
\usepackage{textcomp}
\usepackage{multicol}
\usepackage{multirow}
\usepackage[table]{xcolor}% http://ctan.org/pkg/xcolor
\usepackage{pgfplots} % package used to implement the plot  
\pgfplotsset{width=6.5cm, compat=1.5} 
\usepackage{color,soul}
\usepackage{subcaption}
\usepackage{orcidlink}
\usepackage{pifont}
\usepackage{amsmath}
\usepackage{empheq}
\usepackage{eqparbox}
\usepackage{textcomp}

\setlist{leftmargin=2mm}

\usepackage[a4paper, total={184mm,239mm}]{geometry}
\def\BibTeX{{\rm B\kern-.05em{\sc i\kern-.025em b}\kern-.08em
T\kern-.1667em\lower.7ex\hbox{E}\kern-.125emX}}

\usepackage{multirow}

\usepackage{float}

\def\authnotes{1}
\newcommand{\ayesha}[1]{\ifnum\authnotes=1 \textcolor{red}{AK: #1} \else \fi}

\begin{document}

%-----------------------------------------------------------
% Paper title
% ----------------------------------------------------------

\title{\texttt{RejSCore}: Rejection Sampling Core for Multivariate-based Public key Cryptography}

\author{Anonymous Submission}

%=====================================
% Authors
%=====================================
\author{Malik~Imran$^\dagger$,
       Safiullah~Khan$^\ddagger$,
       Zain~Ul~Abideen$^*$,
       Ciara~Rafferty$^\dagger$,
       Ayesha~Khalid$^\dagger$, \\
       Muhammad~Rashid$^\S$
       and~Máire~O'Neill$^\dagger$ \\

   $^\dagger$ Centre for Secure Information Technologies, Queen's University, Belfast, Northern Ireland, UK \\
   $^\ddagger$ Department of Computing and Mathematics, Manchester Metropolitan University, Manchester, UK \\
   $^*$ Department of Electrical and Computer Engineering, Carnegie Mellon University, Pittsburgh, PA, USA \\
   $^*$ Department of Electrical and Computer Engineering, University of Idaho, Moscow, ID, USA \\
   $^\S$ Department of Computer and Network Engineering, Umm Al-Qura University, Makkah, KSA \\
   $^\dagger$\{m.imran@qub, c.m.rafferty@qub, a.khalid@qub, m.oneill@ecit.qub\}.ac.uk), $^\ddagger$safiullah.khan@mmu.ac.uk \\
   $^*$zabideen@uidaho.edu, $^\S$mfelahi@uqu.edu.sa

}

\maketitle
\thispagestyle{fancy}

    % ----------------------------------------------------------% Abstract of the paper
    %-----------------------------------------------------------
    \begin{abstract}
    Post-quantum multivariate public key cryptography (MPKC) schemes resist quantum threats but require heavy operations, such as rejection sampling, which challenge resource-limited devices. Prior hardware designs have addressed various aspects of MPKC signature generation. However, rejection sampling remains largely unexplored in such contexts. This paper presents \texttt{RejSCore}, a lightweight hardware accelerator for rejection sampling in post-quantum cryptography. It specifically targets the QR-UOV scheme, which is a prominent candidate under the second-round of the  National Institute of Standards and Technology (NIST) additional digital signature standardization process. The architecture includes an AES-CTR-128-based pseudorandom number generator. Moreover, a lightweight iterative method is employed in rejection sampling, offering reduced resource consumption and area overhead while slightly increasing latency. The performance of \texttt{RejSCore} is comprehensively evaluated on Artix-7 FPGAs and 65\,nm CMOS technology using the Area-Delay Product (ADP) and Power-Delay Product (PDP). On Artix-7 and 65\,nm CMOS, \texttt{RejSCore} achieves an area of 2042 slices and 464{,}866~$\mu m^2$, with operating frequencies of 222\,MHz and 565\,MHz, respectively. Using the QR-UOV parameters for security level I ($q = 127$, $v = 156$, $m = 54$, $l = 3$), the core completes its operation in 8525 clock cycles. The ADP and PDP evaluations confirm \texttt{RejSCore's} suitability for deployment in resource-constrained and security-critical environments.
    \end{abstract}

    %=============================================================
    % Keywords
    %=============================================================
    

    %===========================================
    % Introduction
    %===========================================
    \section{Introduction} \label{sec:introduction}
    
    Quantum computing poses a major threat to classical cryptographic systems~\cite{2025review}. 
    In response, the National Institute of Standards and Technology (NIST) began standardizing post-quantum algorithms, focusing on secure key encapsulation and digital signature schemes.
    Specifically, NIST has completed its first post-quantum cryptography (PQC) standardization contest~\cite{alagic2024status}. This contest finalized key encapsulation and digital signature algorithms. The second competition process is now underway, focusing solely on digital signature schemes. NIST’s ongoing digital signature standardization process includes candidates based on diverse mathematical problems. A highly promising area is multivariate-based public key cryptography (MPKC)~\cite{Dey2023}. This approach relies on the challenge of solving systems of multivariate quadratic equations over finite fields, a problem that is recognized as NP-complete~\cite{garey1990guide}. In its second standardization contest, NIST has promoted several MPKC-based schemes, including MAYO~\cite{MAYO}, SNOVA~\cite{SNOVA}, UOV~\cite{UOV}, and QR-UOV~\cite{QR_UOV}. Each presents unique trade-offs in terms of mathematical structure, signature compactness, and computational load.
    
    Despite MPKC algorithms showing mathematical robustness, they do incorporate several computationally intensive elements~\cite{Furue2025UOVVariants} such as rejection sampling. Before rejection sampling, pseudo-random number generators (PRNGs) are used to generate pseudorandom bytes. The typical PRNGs include AES in counter mode (AES-CTR) or hash-based extendable output functions from the SHA3 family~\cite{Furue2023MinRankAttack}. Rejection sampling checks each byte and excludes invalid ones. The valid bytes are then used to form matrices. These matrices are essential for generating and verifying digital signatures. AES-CTR is preferred over hash-based PRNGs in MPKC schemes due to its efficiency and hardware compatibility. Therefore, implementing rejection sampling using AES-CTR in hardware is required for real-time performance and efficient selection of valid bytes. 
    
    Among existing MPKC schemes, MAYO remains the only one that has been addressed through hardware implementations, including field programmable gate arrays (FPGAs) and application-specific integrated circuits (ASICs) platforms~\cite{ha_mayo_fault_tolerant_design, mayo_ms_thesis, whipping_mayo_sujoy}. On the other hand, schemes such as SNOVA, UOV, and QR-UOV have yet to be explored in full hardware contexts. Although a finite field multiplier architecture for UOV and QR-UOV has been proposed~\cite{yasir_mult_QR_UOV}, this work overlooks the implementation of the critical components, such as matrix multiplication and sampling units. Therefore, there is a need for a thorough evaluation of MPKC schemes, particularly rejection sampling, across diverse hardware platforms. 
    As MPKC schemes offer distinct trade-offs involving mathematical structure, signature compactness, and computational complexity, QR-UOV stands out in this context due to its algebraic design. It is well-suited for constrained platforms like embedded systems and edge devices~\cite{Furue2025UOVVariants}. 
    
    This work presents the \textit{first} FPGA and ASIC-based hardware accelerator targeting the rejection sampling block of the MPKC  signature scheme. Our contributions are summarized as:

    \begin{itemize}
        \item A novel hardware-accelerated rejection sampling core, referred to as \texttt{RejSCore}, has been presented. It is tailored specifically for the QR-UOV signature scheme.
    
        \item An efficient architecture for an AES-CTR-128-based PRNG is implemented, comprising (i) an AES-CTR wrapper and (ii) a fully unrolled, pipelined AES-128 encryption engine.
        
        \item A lightweight iterative architecture with an efficient data buffering strategy is proposed for implementing rejection sampling. While this approach introduces an increase in latency, it significantly reduces area overhead.
        
        \item The performance of \texttt{RejSCore} has been comprehensively evaluated on Artix-7 FPGA and 65\,nm ASIC technologies. Figures-of-merit (FoMs) in Area-Delay Product (ADP) and Power-Delay Product (PDP) were used for benchmarking. 
        
    \end{itemize}
    
    The architecture supports QR-UOV at NIST security~level~I and is functionally validated using the public C/C++ reference code~\cite{QR_UOV_REF_CODE}. On Artix-7 FPGA and 65 nm ASIC platforms, \texttt{RejSCore} operates on 222 MHz and 565 MHz post-synthesis frequencies, with area usage of 2042 slices (FPGA) and 464,866~$\mu m^2$ (ASIC). While designed for QR-UOV, its configurable framework allows adaptation to other MPKC schemes by adjusting algebraic constraints and sampler parameters.

    %==========================================
    % Section: Background
    %==========================================
    \section{Background} \label{sec:background}
    
    %==========================================
    % Section: Notations
    %==========================================
    \subsection{Notations}
    
    \begin{tabular}{ll}
        $q$ & prime modulus \\
        $\mathbb{F}_q$ & finite field with $q$ elements for a prime power $q$ \\
        % $\mathbb{B}$ & set $\{0, \dots, 255\}$ of integers represented by a byte \\
    
        $\tau$ & length of random byte string \\
        $\ell, V, M$ & positive integers \\
        $v$ & number of vinegar variables: $v=\ell \cdot V$ \\  
        $m$ & number of oil variables: $m=\ell \cdot M$ \\ 
    
        $\lambda$ & security parameter

    \end{tabular}

    %==========================================
    % Section: Multivariate-based Public key Cryptography
    %==========================================
    
    \subsection{Multivariate-based Public key Cryptography}
    It is a post-quantum technique built on solving multivariate quadratic equations over finite fields~\cite{Dey2023}. One of the critical components is rejection sampling, which filters random inputs to meet algebraic constraints. It exists in two forms: iterative, where each byte is checked individually, and batch, which processes multiple bytes together. AES-CTR and SHAKE-128/256 are common pseudorandom sources. Among many MPKC schemes, QR-UOV stands out for its compact structure and resistance to algebraic attacks~\cite{Furue2025UOVVariants}. 

    %==========================================
    % Section: QR-UOV Preliminaries
    %==========================================
    \subsection{QR-UOV}
    
    The QR-UOV scheme was initially proposed in 2021~\cite{initially_proposed_QR_UOV}. It is a quotient ring-based variant of the Unbalanced Oil and Vinegar signature scheme, i.e., UOV. More precisely, the QR-UOV scheme uses an injective ring homomorphism from the quotient ring $\mathbb{F}_q[x]/(f)$ to the matrix ring $\mathbb{F}_q^{\ell \times \ell}$, where $f \in \mathbb{F}_q[x]$ is a polynomial with $\deg f = \ell$. For complete mathematical constructions, readers are referred to ~\cite{QR_UOV}.

    %==========================================
    % Section: PRNG
    %==========================================
    \subsection{Pseudorandom Number Generator} \label{subsec:PRNG_background}
    
    The \texttt{RejSampPRG} (Algorithm~\ref{alg:algo_rejsampPRG}) serves as the core pseudorandom sampling module in the QR-UOV scheme. The proposed \texttt{RejSCore} hardware architecture directly implements this functionality. QR-UOV supports multiple pseudorandom number generators depending on the security level. For NIST levels I, III, and V, AES in counter mode is applicable using AES-128, AES-192, and AES-256, respectively. Alternatively, SHAKE-128 can be employed for SL-I, while SHAKE-256 is suited for SL-III and SL-V. In this work, AES-128 in counter mode is adopted to accelerate the rejection sampling process for SL-I, primarily due to its hardware efficiency, deterministic behavior, and compatibility with cryptographic accelerators.

    %==========================================
    % Algorithm 1: PRNG
    %==========================================
    
    \begin{algorithm}[htb!]
    \small
    \caption{\texttt{RejSampPRG} procedure~\cite{QR_UOV}}
    \label{alg:algo_rejsampPRG}
    \SetKwInput{KwInput}{Input}
    \SetKwInput{KwOutput}{Output}
    \DontPrintSemicolon
    
    \KwInput{\texttt{seed} $\in \{0,1\}^\lambda$, counter $i$, byte length $\tau$, vector length $n'$}
    \KwOutput{Pseudorandom sequence $(v_1, \dots, v_{n'}) \in \mathbb{F}_q^{n'}$}
    
    $(r_1, \dots, r_\tau) \gets$ \texttt{PRG} (\texttt{seed}, $i$, $8\tau$) \tcp*[r]{Generate $8\tau$ bits of pseudorandomness}
    
    $(v_1, \dots, v_{n'}) \gets$ \texttt{RejSamp} $(r_1, \dots, r_\tau)$, $\tau$, $n'$) \tcp*[r]{Sampling step}
    
    \Return $(v_1, \dots, v_{n'})$
    \end{algorithm}

    In Algorithm~\ref{alg:algo_rejsampPRG}, the \texttt{RejSampPRG} procedure generates a pseudorandom vector $(v_1, \dots, v_{n'}) \in \mathbb{F}_q^{n'}$ from a given seed. It takes four inputs: a seed $\texttt{seed} \in \{0,1\}^\lambda$, a counter $i$, the byte length $\tau$, and the output length $n'$.  Line~1 expands the \texttt{seed} using the \texttt{PRG(seed, $i$, $8\tau$)} function. This produces $\tau$ pseudorandom bytes based on the seed and counter $i$.  Line~2 applies the \texttt{RejSamp} function. It filters and maps these bytes to valid elements in $\mathbb{F}_q$. This process continues until $n'$ valid elements are obtained. The details of the \texttt{RejSamp} procedure will be explained in section~\ref{subsec:RejSamp_background}.

    %==========================================
    % Section: Rejection Sampling
    %==========================================
    
    \subsection{Rejection Sampling}\label{subsec:RejSamp_background}

    The \texttt{RejSamp} procedure is illustrated in Algorithm~\ref{alg:algo_rejsamp}. It accepts a pseudorandom byte string $(r_1, \dots, r_\tau) \in \mathbb{B}^\tau$,  the byte length $\tau$, and a target output length $n'$. The bytes are converted into field elements. Each element must lie in the set $\{0, \dots, q-1\}$, where $q$ is a prime power.

    %==========================================
    % Algorithm 2: RejSamp
    %==========================================
    
    \begin{algorithm}[htbp!]
    \small
    \caption{\texttt{RejSamp}~\cite{QR_UOV}}
    \label{alg:algo_rejsamp}
    \SetKwInput{KwInput}{Input}
    \SetKwInput{KwOutput}{Output}
    \DontPrintSemicolon
    
    \KwInput{Byte string $(r_1, \dots, r_\tau) \in \mathbb{B}^\tau$, byte length $\tau$, vector length $n'$}
    \KwOutput{Vector $(v_1, \dots, v_{n'}) \in \mathbb{F}_q^{n'}$}
    
    \For{$j \gets 1$ \KwTo $\tau$}{
      $v_j \gets$ \texttt{BitsToInteger}(\texttt{BytesToBits}($r_j$) $\land$ \texttt{IntegerToBits}($q$, $8$), $\log_2(q+1)$) \tcp*[r]{Assuming $q$ is Mersenne}
    }
    
    $k \gets n'+1$ \;
    
    \While{$v_k = q$ \textbf{and} $k < \tau + 1$}{
      $k \gets k + 1$ \;
    }
    
    \For{$j \gets 1$ \KwTo $n'$}{
      \If{$v_j = q$}{
        \If{$k < \tau + 1$}{
          $v_j \gets v_k$ \; 
          $k \gets k + 1$ \;
    
          \While{$v_k = q$ \textbf{and} $k < \tau + 1$}{
            $k \gets k + 1$ \;
          }
        }
        \Else{
          $v_j \gets 0$ \;
        }
      }
    }
    
    \Return $(v_1, \dots, v_{n'})$ \tcp*[r]{$v_j \in \{0, \dots, q-1\} \subset \mathbb{F}_q$}
    \end{algorithm}

    \begin{itemize}
        \item Lines 1--2 of Algorithm~\ref{alg:algo_rejsamp} convert each byte $r_j$ into an integer $v_j$, using a bitwise AND between \texttt{BytesToBits}$(r_j)$ and $\texttt{IntegerToBits}(q, 8)$. Since $q$ is a Mersenne prime, this operation extracts the lower $m$ bits of $r_j$. It avoids modular reduction and ensures $v_j \in [0, 2^m)$. 
        
        \item Lines 3--14 handle byte rejection and replacement. A pointer $k = n'+1$ is initialized to search for valid replacements. Lines 4--5 advance $k$ until a valid value is found or the stream ends. 
        
        \item Lines 6--14 iterate over $j = 1$ to $n'$. If $v_j = q$, it is replaced with $v_k$ when available. The algorithm returns the corrected vector $(v_1, \dots, v_{n'})$. All elements are guaranteed to belong to $\mathbb{F}_q$. 
    \end{itemize}

    %===========================================
    % LightWeight Rejection Sampling Core
    %===========================================
    
    \section{Proposed Architecture for Sampling Rejection} \label{sec:prop_core}
    
\begin{figure}[tb]
    \centering
    \includegraphics[width=\columnwidth]{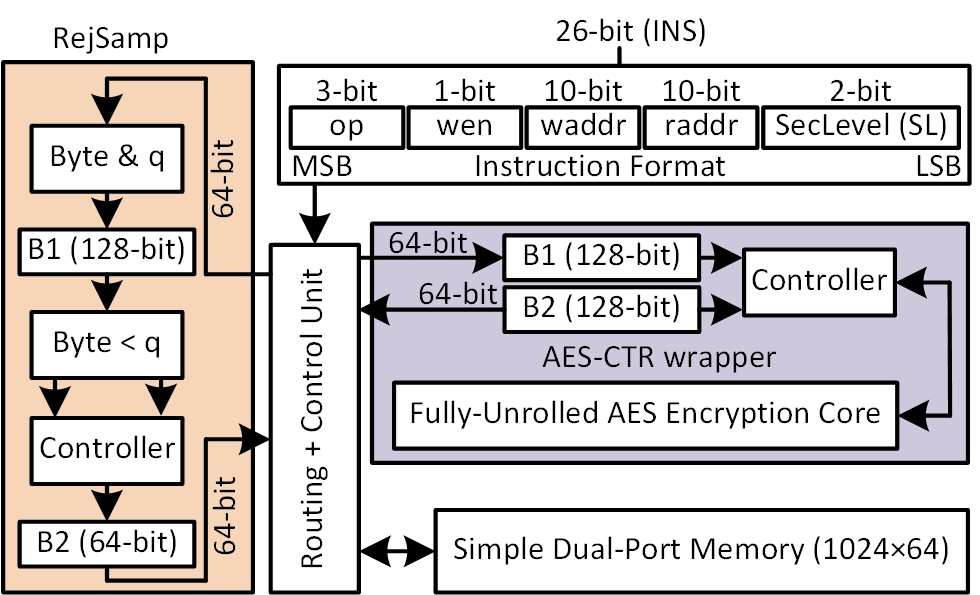}
    \centering
    \caption{\texttt{RejSCore}: Lightweight rejection sampling core for QR-UOV using AES-CTR-based PRNG.
    }
    \label{fig:RejSCore}
    \end{figure}

    We adopt a coprocessor-style architecture for the proposed \texttt{RejSCore}, as shown in Fig.~\ref{fig:RejSCore}. All internal operations between various functional blocks are performed over a 64-bit data path. The architecture consists of five components: (i) a 26-bit instruction set format for operation decoding and configuration, (ii) a dual-port memory unit for pseudorandom data buffering, (iii) a centralized control unit, (iv) an AES-CTR wrapper featuring a fully unrolled AES-128 encryption core, and (v) the RejSamp unit, which implements lightweight iterative rejection sampling logic. Each of these components is elaborated in sections~\ref{subsec:inst_set_format} through~\ref{subsec:RejSamp}.
    
    %%=====================================%%
    %% Fig: NTT-Computecore
    %%=====================================%%
    
    The coprocessor-style architecture in \texttt{RejSCore} targets efficiency and modularity at the same time. It handles pseudorandom generation and rejection sampling independently. A control unit interprets instructions and manages tasks locally. Memory access is fast due to the dual-port design. Modular blocks allow easy upgrades, like new PRNGs or security levels. It also supports embedding countermeasures without changing the entire system. This setup is ideal for secure, low-power embedded platforms.
    
    %%=====================================%%
    %% Section: Instruction Set Format
    %%=====================================%%
    
    \subsection{Instruction Set Format} \label{subsec:inst_set_format}

    As shown in Fig.~\ref{fig:RejSCore}, the instruction set format uses a 26-bit word, denoted as \texttt{INS}. This word is decoded and passed to the control unit to activate the \texttt{RejSCore}.   The decoding process starts from the least significant bit (LSB) and moves to the most significant bit (MSB). The \texttt{INS} contains several fields. A 2-bit field, \texttt{SecLevel} (or \texttt{SL}), selects the required security level. A 10-bit field, \texttt{raddr}, retrieves data from memory after computation. Another 10-bit field, \texttt{waddr}, preloads the \texttt{seed} into the memory unit.  There is also a 1-bit signal, \texttt{wen}, that enables memory writes. Finally, a 3-bit operation code, \texttt{op}, defines the function to execute. It can trigger either \texttt{RejSampPRG} or \texttt{RejSamp}.

    It is important to note that \texttt{RejSCore} currently supports only SL-I. To provide support for other security levels, i.e., SL-III and SL-V, no modifications to the instruction set format are required, as the \texttt{SecLevel} signal is 2 bits wide and can represent up to four distinct options. Likewise, the 3-bit \texttt{op} field allows for up to eight different operations, providing sufficient flexibility to accommodate future extensions without modifying the instruction format of \texttt{RejSCore}.
    
    %%=====================================%%
    %% Section: Memory Unit
    %%=====================================%%
    
    \subsection{Memory Unit}
    \label{subsec:mem_unit}
    
    % \textbf{Memory Unit:} 
    
    The memory unit in Fig.~\ref{fig:RejSCore} acts as a storage element. Its size is determined by the QR-UOV parameters. For SL-I, the relevant parameters are $\ell = 3$, $V = 52$, $M = 18$, and $\tau_2 = 2916$. Here, $\tau_2$ represents the length of the pseudorandom byte sequence generated by Algorithm~\ref{alg:algo_rejsampPRG}. The term $\ell VM = 2808$ corresponds to the number of valid field elements obtained after rejection sampling, as described in Algorithm~\ref{alg:algo_rejsamp} and section~\ref{subsec:RejSamp}.  
    Packing one byte per memory address would require 2916 or 2808 individual addresses for storing $\tau_2$ and $\ell VM$ bytes, respectively. This approach severely impacts the performance of the PRNG accelerator. For example, the AES-CTR core generates 16 bytes of output per cycle. If only one byte is stored per address per cycle, then writing a single AES block would take 16 cycles. This leads to waiting delays and increases the number of required clock cycles. Hence, single-byte packing is inefficient for long sequences.
    
    To address this, we adopt a packing strategy that stores 8 bytes per memory address (i.e., 64-bit words), which reduces both the address space and memory access overhead. With this approach, only 365 addresses are needed for $\tau_2$ and 351 addresses for $\ell VM$, significantly improving cycle counts and memory efficiency.  Applying the same approach to the SL-III parameters ($\ell = 3$, $V = 76$, $M = 26$, $\tau_2 = 6123$) and SL-V parameters ($\ell = 3$, $V = 102$, $M = 35$, $\tau_2 = 11018$), we require 766 and 741 addresses for SL-III, and 1378 and 1339 addresses for SL-V, respectively. In our \texttt{RejSCore} architecture, we use a memory of size $1024 \times 64$ bits, which supports both SL-I and SL-III. For SL-V, the memory must be extended to at least access the 1378 addresses. For FPGA implementation, a dual-port block RAM IP of size $1024 \times 64$ is instantiated to support concurrent read and write operations. In contrast, \texttt{RejSCore} instantiates two distributed memory blocks of size $1024 \times 32$ for ASIC, due to restricted access to commercial memory compilers.

    %%=====================================%%
    %% Section: Control Unit
    %%=====================================%%
    
    \subsection{Routing + Control Unit}
    \label{subsec:control_unit}
    
    At the center of the \texttt{RejSCore} is the controller, which interprets instructions and governs data movement between the AES-CTR wrapper, memory, and RejSamp unit. Upon decoding the 26-bit instruction, the controller configures the target operation, asserts control signals (e.g., \texttt{wen}), and activates relevant functional unit(s). It monitors the \texttt{done} and \texttt{valid} flags from each functional block to manage synchronization and ensure that operations complete correctly. Additionally, it facilitates address management and sequencing for reads/writes across the utilized dual-port memory.
    
    %%=====================================%%
    %% Section: AES-CTR-wrapper
    %%=====================================%%
    
    \subsection{AES-CTR Wrapper}\label{subsec:AES_CTR_wrapper}

    The AES-CTR wrapper in \texttt{RejSCore} performs line~1 of Algorithm~\ref{alg:algo_rejsampPRG}. In this step, \texttt{PRG(seed, $i$, $8\tau$)} expands a seed and counter to generate $\tau$ pseudorandom bytes.  As shown in Fig.~\ref{fig:RejSCore}, the wrapper contains two 128-bit buffers: \texttt{B1} and \texttt{B2}. The \texttt{B1} stores a 16-byte seed. This implementation currently supports only security level SL-I (AES-128 encryption core). However, the design is extensible to higher security levels, SL-III and SL-V, by replacing the AES-128 core with AES-192 or AES-256, respectively. The only required modification is resizing the \texttt{B1} buffer to 192 bits for SL-III and 256 bits for SL-V.  The \texttt{B2} buffer holds the 128-bit output generated by the AES core. After encryption, the bytes in \texttt{B2} are split into two 64-bit chunks. Each chunk is written separately to the memory unit over two clock cycles. This process repeats until all $8\tau$ bits (i.e., $\tau$ bytes) are generated and stored.

    An internal controller/finite state machine (FSM) within the AES-CTR wrapper manages the control flow. It constructs the AES input block using a 64-bit fixed nonce and a 64-bit counter. This counter is created by concatenating a 2-byte initialization vector (IV) from QR-UOV with six zero bytes. Once prepared, the controller activates the AES core and increments the counter for each block. The AES-128 core in \texttt{RejSCore} uses a fully unrolled, 2-stage pipeline design. It executes ten encryption rounds—nine standard rounds and one final round. Each round is mapped to a separate pipeline stage. A second pipeline handles key expansion in parallel. It generates the round keys simultaneously. Every round transformation (substitute-bytes, shift-rows, mix-columns, and key-expansion) requires two clock cycles. Thus, the total latency from seed input to pseudorandom byte output is fixed at 21 clock cycles.

    %%=====================================%%
    %% Section: RejSamp
    %%=====================================%%
    
    \subsection{Rejection Sampling (\texttt{RejSamp})}\label{subsec:RejSamp}

    The RejSamp block with an efficient data
    buffering strategy in Fig.~\ref{fig:RejSCore} implements Algorithm~\ref{alg:algo_rejsamp}. As outlined earlier in section~\ref{sec:background}, it performs two key tasks. First, it generates elements over the finite field $\mathbb{F}_q$. Second, it constructs a vector over $\mathbb{F}_q^m$.  For the first step, the \texttt{Byte \& q} block receives a 64-bit input word containing 8 bytes. It applies a bitwise AND between each byte and the modulus $q$ in parallel. This operation masks the upper bits to keep the values in a valid range, as discussed in section~\ref{subsec:RejSamp_background}. The resulting 8 bytes are stored in buffer \texttt{B1}. To fill the 128-bit internal shift register, the RejSamp unit loads two 64-bit words over two clock cycles.  Next, the bytes in \texttt{B1} are sent to the \texttt{Byte $<$ q} block for validation. This block checks all 16 bytes in parallel against the modulus $q$ and generates two outputs: the candidate bytes and their validity flags. The FSM controller reads these flags and selects only those bytes that satisfy the condition \texttt{byte}~$<$~$q$. It keeps collecting valid bytes until 8 are available. Once ready, it groups these bytes into a 64-bit word, stores them in buffer \texttt{B2}, and writes them to memory in one cycle. This process continues until the required output length $\ell VM$ is reached.

    %============================================
    % Implementation Results and Discussions
    %============================================
    \section{Implementation Results and Discussions} \label{sec:results}
    
    The \texttt{RejSCore} architecture is evaluated on both FPGA and ASIC platforms to demonstrate its efficiency across hardware domains. A direct comparison with existing designs is infeasible, as \texttt{RejSCore} represents the first hardware accelerator for rejection sampling in the QR-UOV scheme.

    %============================================
    % Section: FPGA Evaluation
    %============================================
    
    \subsection{FPGA-based Implementations + Evaluations}\label{subsec:eval_FPGA}
    
    Table~\ref{tab:fpga} presents FPGA implementation results, where column one shows the implemented \texttt{RejSCore} design along with the instantiated functional units. Columns two to four demonstrate hardware area in slices, look-up tables (LUTs) and flip-flops (FFs). The clock cycles and circuit operating frequency are given in columns five and six, respectively. Finally, the last column provides the consumed power. Moreover, the design is synthesized using \texttt{Vivado IDE 2023.2}, targeting an \texttt{Artix-7 XC7A100TCSG324-3} device. The reported results are after the post-place-and-route (PPnR). 

    %%=====================================
    %% Table for FPGA Results
    %%=====================================
    \begingroup
    \setlength{\tabcolsep}{2pt} % Slightly narrower columns
    \renewcommand{\arraystretch}{1.05} % Reduced row spacing for compact layout
    \begin{table}[htbp!] % Changed from table* to table for single-column format
    \footnotesize \centering
    \caption{Implementation Results on Artix-7 FPGA.}
    \label{tab:fpga}
    \begin{tabular}{|p{2.7cm}|p{0.8cm}|p{0.8cm}|p{0.8cm}|p{0.9cm}|p{0.9cm}|p{0.9cm}|}
    \hline
    \textbf{Design Blocks} & \textbf{Slices} & \textbf{LUTs} & \textbf{FFs} & \textbf{Clock Cycles} & \textbf{Freq. (MHz)} & \textbf{Power (W)} \\ \hline
    
    {\texttt{RejSCore}} & 2042 & 5108 & 5130 & 8525 & \multirow{4}{*}{222} & \multirow{4}{*}{1.2} \\
    
    {\hspace{3pt}$\lceil$ AES-CTR wrapper} & 1980 & 4934 & 4911 & 4632 & & \\
    
    {\hspace{9pt}$\lceil$ AES-128-core} & 1708 & 4588 & 4411 & -- & & \\
    
    {\hspace{3pt}$\lceil$ RejSamp} & 117 & 175 & 219 & 3893 & & \\ \hline
    
    \end{tabular}
    \end{table}
    \endgroup

    \subsubsection{Area Evaluation} In terms of hardware resource utilization, the complete \texttt{RejSCore} occupies 2042 slices, utilizing 5108 LUTs and 5130 FFs. The majority of the resource usage stems from the AES-CTR wrapper, which integrates the AES-128 core for cryptographic computations and the RejSamp unit for statistical filtering. As expected, the AES-128 core constitutes the densest logic, consuming 1708 slices, 4588 LUTs, and 4411 FFs, reflecting the complexity of cryptographic operations. The block-RAM (BRAM) usage is not reported in Table~\ref{tab:fpga}. However, \texttt{RejSCore} also utilizes an IP of a BRAM.   
    
    \subsubsection{Power Evaluation} Power analysis using Vivado’s vectorless estimation reports a total on-chip power consumption of 1.212\,W as listed in Table~\ref{tab:fpga}, with 93\% dynamic and 7\% static components. The breakdown shows that BRAMs contribute the largest share at 0.623\,W (56\%), followed by logic (0.166\,W) and clocks/signals. The static device power is reported as 0.091\,W, while the junction temperature remains within safe limits at 30.5$^\circ$C under an ambient of 25$^\circ$C. These results confirm the design’s suitability for medium-power embedded platforms, while still enabling high-throughput post-quantum computations.
    
    \subsubsection{Performance Evaluation} To evaluate the performance of the \texttt{RejSCore} on the targeted Artix-7 FPGA, we used the main parameters of QR-UOV specified in~\cite{QR_UOV} for security level SL-I, with $q = 127$, $v = 156$, $m = 54$, and $l = 3$. The design operates at a post-synthesis frequency of 222\,{MHz}. Using the reported cycle counts, the computation latency of the full \texttt{RejSCore} is approximately:
    \[
    \text{Time} = \frac{\text{Clock Cycles}}{\text{Frequency}} = \frac{8525}{222 \times 10^6} \approx \textbf{38.4}~\mu s.
    \]
    Likewise, the RejSamp requires 3893 cycles ($\approx$ \textbf{17.5}~$\mu s$), and the AES-CTR wrapper alone consumes 4632 cycles ($\approx$ \textbf{20.9}~$\mu s$). These figures demonstrate the suitability of   \texttt{RejSCore} for latency-sensitive applications.
    
    %============================================
    % Section: ASIC Evaluation
    %============================================
    
    \subsection{ASIC-based Implementations + Evaluations}\label{subsec:eval_ASIC}
    
    Table~\ref{tab:asic} presents the logic synthesis results of \texttt{RejSCore} over commercial 65\,nm CMOS technology. The first column provides the implemented design, while the obtained operating frequency (in MHz) is reported in the second column. The total utilized area in $\mu m^2$ is shown in the third column. Similarly, the fourth column reports the standalone area of the instantiated memory unit. The sequential and combinational cells utilization is reported in the fifth and sixth columns, respectively. The last two columns report the static and dynamic power consumption of the \texttt{RejSCore}. Moreover, the synthesis is performed using the \texttt{Cadence Genus} tool, targeting the typical process-voltage-temperature (PVT) corner with a low-power flavor. 

    %============================================
    % Table: ASIC Results + Evaluation
    %============================================
    
    \begingroup
    \setlength{\tabcolsep}{2pt} % Slightly narrower column spacing
    \renewcommand{\arraystretch}{1.05} % Compact row spacing
    \begin{table}[htbp!] % Changed from table* to table for one-column layout
    \footnotesize \centering
    \caption{Implementation Results on 65\,nm ASIC.}
    \label{tab:asic}
    \begin{tabular}{|p{1.5cm}|p{0.9cm}|p{1.0cm}|p{0.8cm}|p{0.65cm}|p{0.9cm}|p{0.75cm}|p{1.1cm}|}
    \hline 
    \textbf{Design} & \textbf{Freq. (MHz)} & \textbf{Total Area} ($\mu m^2$) & \textbf{Mem. Area} ($\mu m^2$) & \textbf{Seq. Cells} & \textbf{Comb. Cells} & \textbf{Static Power (mW)} & \textbf{Dynamic Power (mW)} \\ \hline
    \texttt{RejSCore} & 565 & 464,866 & 71471 & 7348 & 162661 & 0.005 & 0.124 \\ \hline
    \end{tabular}
    \end{table}
    \endgroup

    \subsubsection{Area Evaluation} The total occupied area of \texttt{RejSCore} after the synthesis and place-and-route steps is 464,866~$\mu m^2$, out of which 71471~$\mu m^2$ is attributed to the dual-port memory blocks used to store a 16-byte seed. As discussed earlier in section~\ref{subsec:mem_unit}, for ASIC, the memory of size $1024\times 64$ is partitioned into two independent 1024$\times$32-bit SRAM instances. This distributed memory architecture enhances parallel access and provides better performance and robustness than a single, monolithic memory would allow. Moreover, \texttt{RejSCore} can operate at a target frequency of 565~MHz, demonstrating the high-performance capability under realistic timing constraints.
    
    \subsubsection{Power Evaluation} Power analysis reveals an extremely low static power consumption of just 0.005\,mW, highlighting the energy efficiency of the design when idle. The dynamic power consumption under switching activity is reported as 0.124\,mW, which remains well within acceptable bounds for low-power cryptographic accelerators, especially when targeting energy-constrained systems.
    
    \subsubsection{Performance Evaluation} We used the same approach and QR-UOV parameters as used in section~\ref{subsec:eval_FPGA} (for FPGA) evaluation to assess the performance of the \texttt{RejSCore} on a commercial 65\,nm technology. Using the reported cycle counts, the computation latency of the full \texttt{RejSCore} is approximately:
    \[
    \text{Time} = \frac{\text{Clock Cycles}}{\text{Frequency}} = \frac{8525}{565 \times 10^6} \approx \textbf{15.0}~\mu s.
    \]
    Similarly, the RejSamp unit requires 3893 cycles ($\approx$ \textbf{6.8}~$\mu s$), and the AES-CTR wrapper alone consumes 4632 cycles ($\approx$ \textbf{8.1}~$\mu s$).

    %============================================
    % Section: FoMs
    %============================================
    
    \subsection{FoMs Analysis Across FPGA and ASIC Platforms}\label{subsec:fom_compare}
    
    To evaluate the efficiency of the \texttt{RejSCore} design on FPGA and ASIC platforms, we use two standard FoMs: (i) ADP and (ii) PDP. These metrics, summarized in Table~\ref{tab:adp_pdp}, capture trade-offs between hardware resource usage, performance, and energy efficiency by incorporating the critical path delay (CPD) into the analysis:
    
    \[
    \text{ADP}_{\text{ASIC}} = \text{Area}_{\mu m^2} \times \text{CPD}_{\text{(s)}} \quad [\mu m^2 \cdot s]
    \]
    \[
    \text{ADP}_{\text{FPGA}} = \text{LUTs} \times \text{CPD}_{\text{(s)}} \quad [\text{LUT} \cdot s]
    \]
    \[
    \text{PDP} = \text{Total Power}_{\text{(mW)}} \times \text{CPD}_{\text{(s)}} \quad [\text{mW} \cdot s]
    \]
    
    %============================================
    % Table: FoMs
    %============================================
    
    \begingroup
    \setlength{\tabcolsep}{2pt} % Slightly tighter column spacing
    \renewcommand{\arraystretch}{1.05} % Slightly reduced row spacing for compactness
    \begin{table}[htbp!] % Changed from table* to table for single-column layout
    \footnotesize \centering
    \caption{CPD, ADP, \& PDP Comparison Across Platforms.}
    \label{tab:adp_pdp}
    \begin{tabular}{|p{3.0cm}|p{0.9cm}|p{2.6cm}|p{1.75cm}|}
    \hline 
    \textbf{Platform} & \textbf{CPD (ns)} & \textbf{ADP ($\mu m^2 \cdot s$)} & \textbf{PDP (mW$\cdot$s)} \\ \hline
    
    ASIC (65\,nm) & 1.77 & $8.23 \times 10^{-4}$ & $2.28 \times 10^{-10}$ \\ \hline
    FPGA (Artix-7) & 4.50 & $2.30 \times 10^{-5}$ LUT$\cdot$s & $5.40 \times 10^{-9}$ \\ \hline
    FPGA (Tech-scaled)$^\dagger$ & 4.50 & $1.24 \times 10^{-4}$ & $5.40 \times 10^{-9}$ \\ \hline
    
    \end{tabular}
    
    \raggedright
    \footnotesize{$^\dagger$LUT area scaled from 28\,nm to 65\,nm using $\text{Area}_{65} = \text{Area}_{28} \times \left( \frac{65}{28} \right)^2$.}
    \end{table}
    \endgroup
    
    The FPGA implementation is based on Xilinx Artix-7, fabricated using a 28\,nm process technology. Since a direct comparison between different technology nodes can be misleading, we introduce a scaling factor to normalize the FPGA's LUT-based area estimate to the 65\,nm CMOS node used in the ASIC flow. This scaling is performed using the classical technology scaling rule:
    \[
    \text{Area}_{65} = \text{Area}_{28} \times \left( \frac{65}{28} \right)^2
    \]
    Applying this transformation allows a normalized ADP estimate for the FPGA, revealing its relative inefficiency due to the overheads of programmable fabrics. 
    
    Table~\ref{tab:adp_pdp} shows that the ASIC implementation achieves significantly lower ADP and PDP than the FPGA, with a shorter CPD of 1.77\,ns compared to 4.5\,ns. Even after area scaling, the FPGA's ADP remains about 6$\times$ higher, reflecting the efficiency gap between custom silicon and programmable logic in performance-critical cryptographic primitives.

    %===============================================
    % Discussion
    %===============================================
    \subsection{Discussions} \label{subsec:discussion}
    
    The FPGA and ASIC results in section~\ref{subsec:eval_FPGA} and section~\ref{subsec:eval_ASIC} confirm the efficiency of the proposed \texttt{RejSCore} architecture for rejection sampling in QR-UOV. On Artix-7 FPGA, the design operates at a post-synthesis frequency of 222~MHz. It occupies 2042 slices and utilizes 5108 LUTs. The entire operation completes in 8525 clock cycles.  In contrast, the ASIC implementation operates on a frequency of 565~MHz. It consumes an area of 464{,}866~$\mu m^2$ and requires only 0.124~mW of dynamic power. These results demonstrate the portability and high performance of \texttt{RejSCore} across platforms.
    
    The FPGA analysis shows that the AES-128 core dominates resource usage with 1708 slices, 4588 LUTs, and 4411 FFs, highlighting its cryptographic load. The RejSamp unit with an iterative buffering approach is lightweight, completing in 3893 cycles. For ASIC, distributed 1024$\times$32 SRAM blocks mitigate compiler constraints while ensuring high throughput. Architectural portability and energy efficiency are evident in PDP and ADP comparisons across platforms. FPGA achieves a PDP of $5.40 \times 10^{-9}$~mW$\cdot$s, while ASIC reaches a notably lower $2.28 \times 10^{-10}$~mW$\cdot$s.

    Although the current design does not incorporate explicit physical attack countermeasures, the \texttt{RejSCore} architecture demonstrates a hardware-controlled and deterministic execution flow, minimizing observable timing variations. This structure is well-suited for integration with side-channel and fault detection mechanisms on both FPGA and ASIC platforms, with minimal area or performance overhead. Thus, \texttt{RejSCore} lays the groundwork for future secure implementations through modular, performance-verified building blocks.

    %===============================================
    % Conclusions
    %===============================================
    \section{Conclusions and Future Directions} \label{sec:conclusion}

    % This work has presented a hardware-accelerated rejection sampling core, \texttt{RejSCore}, for the QR-UOV signature scheme. The core employs an AES-CTR-128 based PRNG (consisting of an AES-CTR wrapper plus a fully unrolled, pipelined AES-128 engine) and a lightweight iterative rejection-sampling architecture that reduces area at the expense of higher latency. Implemented on Artix-7 FPGAs and synthesized in 65\,nm CMOS, \texttt{RejSCore} completes in 8{,}525 cycles for Level I parameters ($q{=}127$, $v{=}156$, $m{=}54$, $l{=}3$). ADP and PDP results indicate suitability for resource-constrained, security-critical deployments. The present work prioritizes performance rather than physical-attack resilience; future work will incorporate SHAKE-based PRNG variants required by QR-UOV, extend coverage to all security levels, and integrate \texttt{RejSCore} into full UOV/QR-UOV accelerators.

    This work has presented a hardware-accelerated rejection sampling core, \texttt{RejSCore}, for the QR-UOV signature scheme. The core employs an AES-CTR-128 based PRNG (consisting of an AES-CTR wrapper plus a fully unrolled, pipelined AES-128 engine) and a lightweight iterative rejection-sampling architecture that reduces area at the expense of higher latency.
    % The core employs an AES-CTR-128-based PRNG, comprising two key components: (i) an AES-CTR wrapper and (ii) a fully unrolled, pipelined AES-128 encryption engine. To efficiently realize the rejection sampling logic, we proposed a lightweight iterative architecture that minimizes area overhead, at the cost of increased latency due to its sequential nature. 
    \texttt{RejSCore} has been implemented on Artix-7 FPGAs and synthesized for a commercial 65\,nm CMOS technology. Using the main QR-UOV parameters for security level I ($q = 127$, $v = 156$, $m = 54$, $l = 3$), the core requires 8525 clock cycles for operation. Performance evaluations in terms of ADP and PDP confirm its suitability for resource-constrained and security-critical applications. The current \texttt{RejSCore} design prioritizes hardware-based performance evaluation and does not address physical attack resilience, which is a key area for future work. Moreover, support for SHAKE-based PRNG variants, as required by QR-UOV, will also be incorporated and requires thorough evaluation alongside the rejection sampling logic. Future extensions include support for all security levels and the integration of \texttt{RejSCore} into full PQC accelerators for UOV and QR-UOV.

     %===============================================
    % Acknowledgements
    %===============================================
    
    \section*{Acknowledgment}
    
    The authors thank the QR-UOV team for their valuable discussions and support.

    %===============================================
    % Acknowledgements
    %===============================================
    
    \section*{Funding}
    
    This work is funded by the Integrated Quantum Networks (IQN) Research Hub (EP/Z533208/1).
    
    \bibliographystyle{ieeetr}
    \bibliography{references}

\end{document}